# DyFraNet: Forecasting and Backcasting Dynamic Fracture Mechanics in Space and Time Using a 2D-to-3D Deep Neural Network

Yu-Chuan Hsu[1,2], Markus J. Buehler[1,2]*

[1] Laboratory for Atomistic and Molecular Mechanics (LAMM), Department of Civil and Environmental Engineering, Massachusetts Institute of Technology, 77 Massachusetts Ave., Cambridge, MA 02139, USA
`
[2] Center for Computational Science and Engineering, Schwarzman College of Computing, Massachusetts Institute of Technology, 77 Massachusetts Ave., Cambridge, MA 02139, USA

*Corresponding author: mbuehler@MIT.EDU, +1.617.452.2750

**Abstract:** The dynamics of materials failure is one of the most critical phenomena in a range of scientific and engineering fields, from healthcare to structural materials to transportation. In this paper we propose a specially designed deep neural network, DyFraNet, which can predict dynamic fracture behaviors by identifying a complete history of fracture propagation – from cracking onset, as a crack grows through the material, modeled as a series of frames evolving over time and dependent on each other. Furthermore, this model can not only forecast future fracture processes but also backcast to elucidate the past fracture history. In this scenario, once provided with the outcome of a fracture event, the model will elucidate past events that led to this state and will predict the future evolution of the failure process. By comparing the predicted results with atomistic-level simulations and theory, we show that DyFraNet can capture dynamic fracture mechanics by accurately predicting how cracks develop over time, including measures such as the crack speed, as well as when cracks become unstable. We use GradCAM to interpret how DyFraNet perceives the relationship between geometric conditions and fracture dynamics and we find DyFraNet pays special attention to the areas around crack tips, which have a critical influence in the early stage of fracture propagation. In later stages, the model pays increased attention to the existing or newly formed damage distribution in the material. The proposed approach offers significant potential to accelerate the exploration of the dynamics in material design against fracture failures and can be beneficially adapted for all kinds of dynamical engineering problems.

1. Introduction

Fracture is a continuous process of how material disintegrate over time, typically due to external loading conditions[1,2,3]. Over the past few decades, much research has been done to study the fundamental fracture mechanisms with simulations to at different scales to determine the relationship between material geometry and toughness[4,5,6,7,8]. However, the dynamics of fracture mechanics remains a challenging topic even with the state-of-art simulation techniques because many physical factors such as mechanical properties, geometric conditions, strain rates or variations in boundary conditions need to be concerned[9]. All these characteristics interactively contribute to stress concentration and redistribution at crack tips, the core mechanism governing crack development[10], so the design space of superior materials against fracture is extremely large and it is not feasible for us to explore all possible design scenarios from the ground up. To overcome this limit, we need a systematic and efficient way to learn and analyze fracture behaviors that can effectively bridge atomistic-level techniques, and rapid screening of phenomena, to link across scales and modalities of exploration.

Conventional ways to model fracture include molecular dynamics[11], FE[12], and phase-field methods[13,14], but these methods tend to be computationally costly as the complexity of the materials grows. Here we pose the question, could an ML method be trained to learn atomistic-level dynamics of fracture, and offer a rapid assessment tool? In our earlier work, we proved that the relationship between geometries and crack patterns can be learned and predicted by a machine learning model[15,16]. Recent advances in applying ML to dynamical problems suggest that deep learning can indeed be a tool to learn and discover time-dependent physical phenomena, such as self-learning conservation laws[17] or predicting dynamical systems[18].

In order to help address dynamic fracture problems based on a generalizable platform, a ML model should be able to provide information in successive periods allowing changes to be made in a system. However, earlier work for time-series prediction typically requires a recursive sequence input to make prediction on the next frames[19,20]. This poses a challenge for realistic engineering problems since we usually do not have any other information rather than initial condition before conducting computational or physical experiments. Also, to complete the whole time-series prediction, these methods need to take the latest predicted output to be the next input iteratively to the targeted timepoint. However, during the next-frame predicting process of a fracture propagation problem, a next-frame predictive model might consider a discrepancy as a crack and then predict another crack next to it, where all errors naturally and inevitably accumulate and eventually will go viral on the predicted frame through these approaches. Lastly, this frame-to-frame approach does not reflect the fact some physical information is passed down from the beginning to the end, such as position-wise stiffness, which would weaken time-dependent features in fracture dynamics. This calls for an approach that considers the full long-range of temporal and spatial relationships that govern complex dynamical failure phenomena.

Dynamic fracture is not only important to understand how engineering structures fail, but also critical to understand how to design better materials. For instance, by learning from existing creatures in nature, recent work has found that unusual geometric arrangements of microstructures in biomaterials can lead to extraordinary mechanical properties to help them survive, especially under dynamical loading conditions[21,22]. For example, most of the seashells developed strong and tough exoskeletons that can be attributed to the hierarchically structured brick-and-mortar arrangement in the microstructure to form different kinds of mechanisms[23,24,25], so a shell could endure more external impact before it breaks into pieces. In addition to seashell animals, bamboo, a plant with countless uses, possesses great toughness against fracture by its functionally graded hierarchical structure in which the fibers in the outer and inner regions have different levels of strength[26,27]. Since organisms in nature only have limited access to materials, the most economical way to mitigate physical threats is optimizing the arrangement materials on their bodies to their best. In the past years, thanks to the findings of structural designs from nature, the development of stronger and tougher materials was making some remarkable progress[28,29,30]; however, the search for superior materials is still strongly relying on a time-consuming research cycle of modeling, simulations, and experiments that often practically limits the design space that can realistically be explored.

To accelerate the exploration of superior materials, a data-driven technique to study dynamic fracture mechanisms is a viable paradigm because it saves computational efforts to rapidly and repeatedly obtain the dynamic properties such as crack patterns or crack speed at the atomistic level that is fundamental to understanding fracture dynamics. The data-based design also allows for a direct integration of synthetic simulation data with experimental data to address a variety of tine-tuned downstream tasks.

How do we reach this goal? In order to reveal dynamic fracture mechanics in materials, we need a model that can understand the relationship between the spatial complexities captured in the microstructure geometries and the temporal history of fracture evolution. We propose a neural network, DyFraNet, based on a physics-inspired 2D-to-3D structure allowing us to introduce time information as a new dimension for the core Long Short-term memory (LSTM) unit[31] to learn fracture dynamics from our dataset, collected from ground-truth atomistic simulation results on crystalline solids in different orientations and stiffnesses.

## 2. Results

First, a computational experiment is designed to study various fracture behaviors affected by material properties, such as stiffness and geometries, to generate a synthetic dataset based on atomistic simulation. As shown in **Fig. 1**, the simulation model is a 2-D rectangular slab, of which the width is 80 and length is 100 units of hexagonal lattice (details of the atomistic simulations, the deep learning model development and training, and other critical details are included in **Materials and Methods**).

We use Lennard-Jones (LJ) 12-6 potential (**Eq. 1**) with 10 different non-dimensional energy coefficients, $\varepsilon$, ranging from 0.1 to 30, which will give us different crack speeds that in theory, are limited to three wave velocities[32]: longitudinal wave speed $c_l = 9\sqrt{\varepsilon/m}$, transverse wave speed $c_t = c_l/\sqrt{3}$, and the Rayleigh wave speed $c_R \approx c_t$, where $m$ is the mass of an atom in the model.

As the LJ energy coefficient becomes larger, the material becomes more rigid so that the crack is predicted to propagate faster. Furthermore, we test these models on 9 different orientation angles to enrich the scope on the effect due to the changes of geometries to observe a wide range of dynamic fracture behaviors.

$$V(r) = 4\varepsilon \left[ \left(\frac{\sigma}{r}\right)^{12} - \left(\frac{\sigma}{r}\right)^{6} \right] \quad \text{Eq. 1}$$

As shown in **Fig. 2(a)**, the trained DyFraNet can either forecast or backcast the credible fracture propagation result depending on the timepoint of the input state, after trained against totally 900 cases of mode I simulations. As can be seen in **Fig. 2(b ,c)**, DyFraNet captures the physical meanings embedded dataset by accurately predicting *when and where* a crack would occur. It is also interesting to see how our model makes a better prediction to provide denser and sharper patterns for backcasting prediction just as if it increases its confidence after seeing the final result to trace back what happened to the material at earlier time points. Most importantly, this shows that a specially designed ML model can perform once-for-all predictions to help address a real engineering problem about how initial conditions could affect fracture dynamics. Moreover, the idea here of using an ML model to do backcasting can be beneficially adapted to many potential applications for post-fracture analysis. Rather than using traditional recursive methods (e.g. MD, or FEM models where we have to step through time), the model framework proposed here can instantly provide a comprehensive picture of all spatial and temporal details.

With this framework, we further depict the overall comparison of the crack growth over time as shown in **Fig. 3(b)**, which follows the similar trend as what we see in the ground truth MD simulations. As can be verified in in **Fig. 3(c, d)**, DyFraNet shows good agreement with MD simulation results specifically regarding the correlation in the different stages between the crack growth rate and the LJ energy coefficients. This clearly shows that our model captures the subtle relationship between the nature of chemical bonds captured via variations in the force fields and the resulting dynamic fracture behaviors, interpreting physical meanings from low-scale physical parameters to higher-scale dynamics. It is also noteworthy that our model implies the likelihood of crack in a wide range of the given spacetime in the late stages of crack development, which is reflecting the fact that the crack development after certain timepoint could have many possibilities and thus becomes inconsistent every time when we simulate them. Our results show that DyFraNet can identify when and where the crack branching phenomena could happen during fracture propagation. Therefore, the predicted history of crack propagation contains essential information of crack lengths and patterns over time that can be taken as an indicator to estimate the dynamic fracture toughness of the current geometry. This can be crucial information for dynamical material design.

At the next step, as shown in **Fig. 4(a)**, we present a scenario of using our approach to solve a *de novo* composite design task to achieve a target dynamical fracture property. In this experiment, we use the same model only trained with single crystals with various stiffness to verify the idea of training a deep learning model with limited yet fundamental knowledge to carry out advanced tasks such as composite design, in realms beyond the types of scenarios seen in the training data. Different patterns of soft material lie in the middle of the hard material to help us examine whether the model can make accurate predictions on the dynamic behaviors that has not been learned. Based on the findings shown in **Fig. 4(b)**, not only can DyFraNet show where the crack would possibly go, but it also can perceive the physical fact that the breakage occurs in the soft material more easily than in harder material and then creates branches out of the main crack. This indicates a great promise for generalization.

We now explore how exactly DyFraNet makes its prediction depending on different temporal-spatial distributions of material damage. Using the Grad-CAM approach[33], as shown in **Fig. 4(c)**, we visualize the activation map of the core layer of the model, that is, the output of the LSTM unit. We find that in the early stage, DyFraNet is focusing on the geometries, for example, whether any soft material is located around, near the crack tip to determine the first and critical step of the crack development. Then, during the primary failure development stage – during which dynamic fracture behaviors become more complex, and when much larger damage areas are involved – the model collects more geometrical information by paying attention to a range of different areas that might contribute to fracture mechanics, simultaneously. Finally, as the crack development slows down and eventually stops in the late stages, the model formulates a connection of all the areas it has identified as damaged areas. It is important to note that due to the special structure to make the prediction all at once, DyFraNet is able to appropriately collect and process all different pieces for each stage and then incorporate them all into a comprehensive understanding, which is ultimately the key to predict the whole failure scenario. This learned feature of the model makes sure that the predicted time-series of physical states by DyFraNet are strongly connected and the dynamic behavior can be accounted for. Hence, with our approach, one can obtain validated predictions of fracture history starting from simple geometries without going through the time-consuming, recursive cycle of modeling, simulations, and even experiments and then focus directly on the design itself to explore the possibilities lying in the design space.

## 3. Discussion and Conclusion

We proposed a new idea to utilize the intrinsic features of an unconventional neural network architecture that consists of a 2-D convolutional encoder, an LSTM, and a 3-D deconvolutional decoder, where the encoder extracts geometric information, the LSTM learns from low-dimensional geometric features and the corresponding behaviors along with time. The decoder part expands the entire time-and-space behaviors for fracture propagation predictions in a variety of physical dimensions. Unlike the other widely used machine learning strategies that typically conduct frame-to-frame recursive predictions (such as ConvLSTM2D networks), our model achieves a much deeper, comprehensive insight into mechanisms and processes; predicting the entire change of microstructural damage over time. This opens up a way to optimize material design for dynamic mechanical properties, as we have demonstrated with a few examples shown in **Fig. 4**.

In addition, the model is trained to capture a range of dynamic fracture behaviors by training it with geometric input at different time points and the associated corresponding fracture history, used as a hint for all possible causal relationships existing in the dataset. That way, the model can solve the task of obtaining results of fracture propagation at any time point. By doing so, we delineate three key applications: (a) taking the input from the beginning before any crack happens and then forecasting what would happen next, (b) taking the input at the end after all cracks happen and backcasting what could happen in the past, and (c) taking the input from any time point to partially forecast and backcast the overall result at the same time. We strongly believe that a model that is capable of all these scenarios can be considered a realistic surrogate model for simulation or even experiments for us to study dynamic fracture mechanics.

*Key results and outlook*

DyFraNet is designed for forecasting and backcasting fracture propagation over time based on a comprehensive understanding of dynamic fracture mechanics in materials, much more efficiently than what humans can do with case-by-case simulation trials and/or similar experimental campaigns. Existing simulation or experiment techniques can help us uncover details of the material behavior under a specific set of conditions, but there exist significant computational and/or experimental challenges to effectively utilize the knowledge to discover a range of new materials better than we already have. Here, our proposed approach has great potential to fill the gap between what we have learned from a set of simulations or experimental campaigns, and amalgamating these insights into a comprehensive predictive model by training a machine learning model to capture underlying physics and provide us with reliable predictions. In the future, with the input of data specialized for desired properties, this approach can accelerate the development of material design for better performance against dynamic failure, where general-purpose models trained from, say MD models, can be fine-tuned against specific experimental data.

The focus of this work was to explore systematic relationships between the potential shape and crack dynamics, and this is best done using a simple interatomic model. The main limitation of the current approach is that the output of DyFraNet is designed to be a 3-D matrix in single-channel only for binary prediction of crack and non-crack areas. However, this can be extended to hold multi-channel values for more complex physical behaviors, for example, using a multi-channel color map to describe stress, strain, or temperature fields, and also a range of different material constituents. Future work could apply the method, either using retraining or using transfer learning with other physical factors, to realistic material models, for example, using potentials beyond LJ, such as EAM for alloys, to better describe complex mechanical behaviors on different kinds of materials.

**4. Materials and Methods**

*4.1 Molecular Dynamics Simulation*
To construct a dataset rich in dynamic fracture mechanisms, we first run mode-I tensile test simulations on different orientations and energy coefficients for LJ potential under the same strain rate[34]. The corresponding stress-strain curves are shown in **Fig. 3(a)**. By looking at how these cracks propagate at certain speeds or along certain patterns in these simulation results, our predictive model can find a way to relate their behaviors to the input geometries. Meanwhile, we can analyze the crack speeds and paths from simulations and our predictive model to evaluate the performance of materials against fracture failure.

As can be seen in **Fig. 5(a)**, the simulation results are first visualized from MD simulations into images in grayscale, in which the black pixels denote the crack regions, and the white ones denote the non-crack. Among these simulation cases, we show that the crack could develop in different speeds and along with different paths given different stiffnesses and orientations. Based on the simulation results, the crack propagates faster in harder material, and the pattern is not only controlled by the stiffness but also by the orientation, which makes it rich in various dynamic mechanisms that can be learned by our model.

*4.2 Image-processing techniques and dataset generation*
As shown in **Fig. 5(b)**, through our image-processing techniques, the simulation results that form the input dataset will be transformed into the data pairs of 2-D images as the input where all values are in 2 channels to handle the geometries and physical factors and 3-D matrices as the output that has single-channel Boolean values to indicate where the crack is located to train our predictive model. It is noted that the black pixels become -1 in our input matrices, distinguished from the chosen physical parameters, such as the coefficients for the pair potential. In this work, since we are focusing on the effects of orientation and the hardness of materials on fracture dynamics, the input images only contain two channels for both, which is flexible for the other targeted properties, such as Young's modulus, bond length, or Poisson's ratio. On the other hand, the output matrices take the true-false criteria to indicate where and when cracks would be located.

The main reason to design a 2-D to 3-D model instead of using a next-frame predictive model, such as ConvLSTM 2D-based neural networks[35], is that this model can make predictions of the whole history of fracture behaviors from a single frame, while the next-frame predictive models must take multiple frames, which is hardly able to replace simulations. Also, during the crack propagation, cracks will remain over time after their occurrence, which can be considered a time-dependent process, so the output can be resampled as a 3-D matrix that provides the time-space information and the causality we want the model to figure out. Therefore, the neural network we develop is a perfect match to help us learn these pairs of geometries and the corresponding crack dynamics over time.

*4.3 LSTM-based 2D-to-3D convolutional neural network*
The integrated model consists of three main components as shown in **Fig. 6(a)**, designed for different purposes. The first component is composed of two 2-D convolutional layers as an encoder to extract and embed geometric features of an input image into a low-level latent space. Then, these low-level features will be transformed by an LSTM unit into spatial-temporal space. A 3-D decoder that is combined with two transpose 3-D convolutional layers will expand the space and time dimension of these features altogether into the desired time and length scale. Finally, the last 3-D convolutional layer will extract the up-sampled features to indicate when and where cracks should happen. Unlike conventional next-frame approaches, with this model, we can predict all frames together, more importantly, with only single-frame input, which allows us to predict fracture behaviors over time based on any time point.

We adopt a GELU activation[36] function for each convolutional layer except for the last output layer, using sigmoid for multiple binary classification of crack or non-crack. Also, since the task represents a binary classification problem, we use binary cross-entropy for the loss and compute the ratio of the number of correct predictions to the total number of voxels in the whole 3D output matrix for accuracy. The learning curve in the training history, as shown in **Fig. 6(b)**, shows that the model learns smoothly and efficiently from the pairs of geometries and the fracture behaviors over time that is rich in dynamic fracture mechanisms.

**Acknowledgements:** We acknowledge support by the Office of Naval Research (N000141612333 and N000141912375), AFOSRMURI (FA9550-15-1-0514) and the Army Research Office (W911NF1920098).

**Conflict of interest**
The authors declare no conflicts of interest.

**Author contribution statement**
MJB designed the main concept of this work and oversaw the project. MJB and YCH designed image processing techniques. YCH developed the deep learning model and carried out the MD simulations for the dataset. Both authors wrote and edited the paper and contributed to the scientific research design and interpretations.

**Data and code availability**
The dataset and the source code that support the findings of this study are available at https://github.com/lamm-mit/DyFraNet. Additional materials are available from the corresponding author, MJB, upon reasonable request.

**Supplementary Information**

- Videos of simulation and predicted results can be found via https://github.com/lamm-mit/DyFraNet

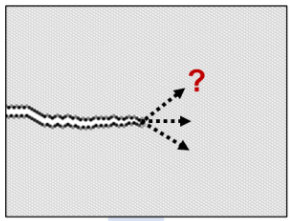
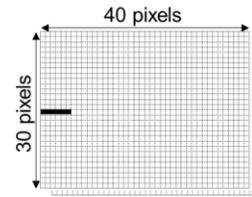
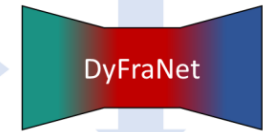
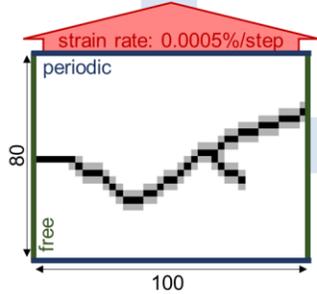
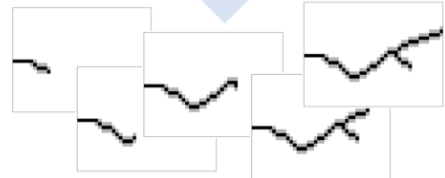

**Fig. 1:** The overall schematic diagram of this research. Here we present a deep learning framework with our specially designed network, DyFraNet, to learn from MD simulation results to study fracture dynamics concerning different kinds of initial conditions. The strategy is to take an image-based structure to include physical factors that we would like to survey as the input, and then let the DP model to learn the relationship between these initial conditions and the overall fracture history altogether at once.

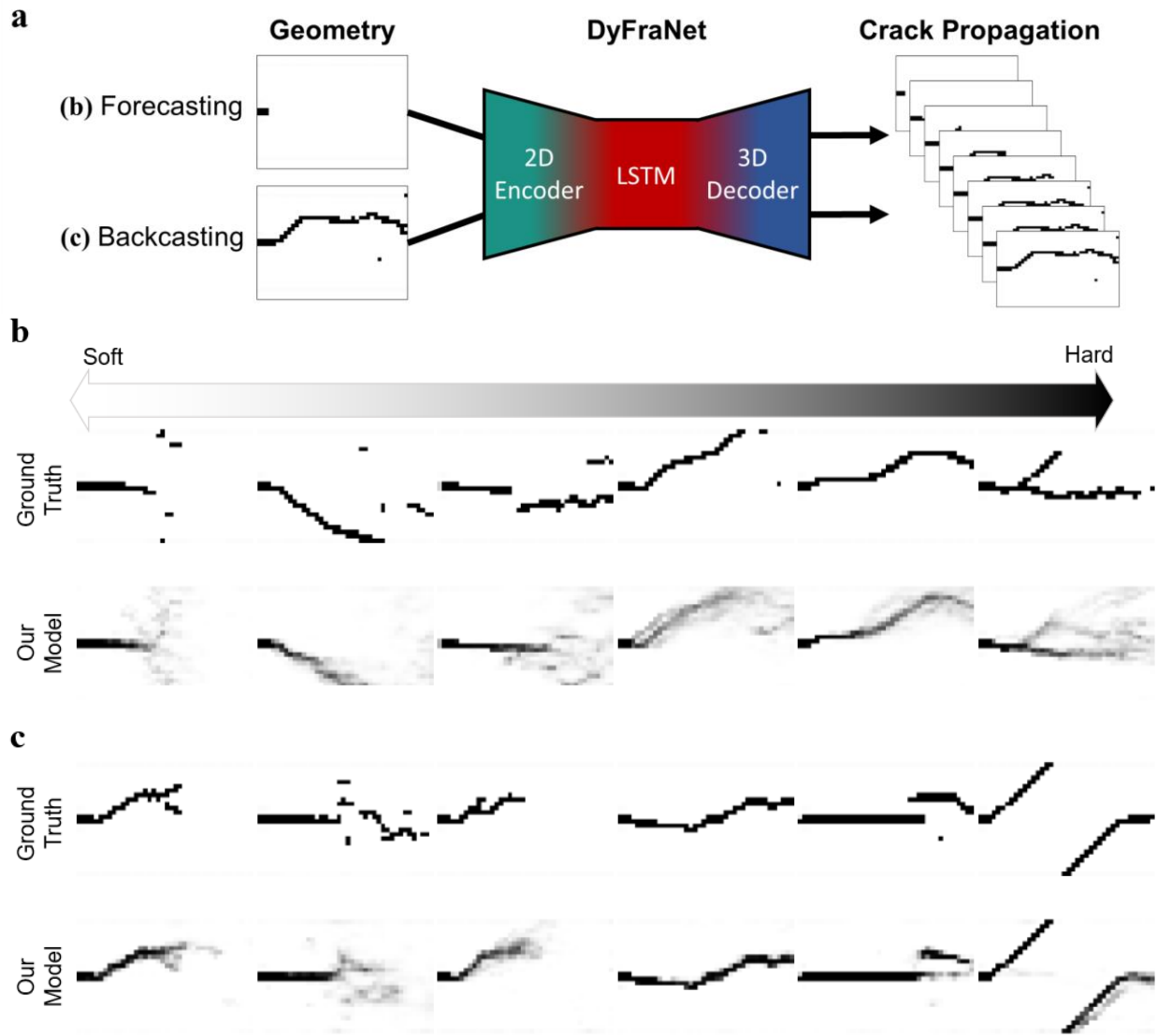

**Fig. 2:** Panel a shows two applications, forecasting and backcasting, of this approach for the model to make the prediction. The predicted histories of fracture propagation by (b) forecasting and (c) backcasting, all frozen at the same time point to show the difference, by our model based on only one single-frame input, in comparison with the ground truth from simulations. (For more details, please see **Movie S1**.) From left to right, we show results for 12 different cases with low to high stiffness. Overall, the cracks branch more easily in softer materials, while the crack speed is faster in harder materials. Both properties can be predicted accurately by our model, which proves our model can capture the subtle relationship between crystalline geometries and fracture dynamics.

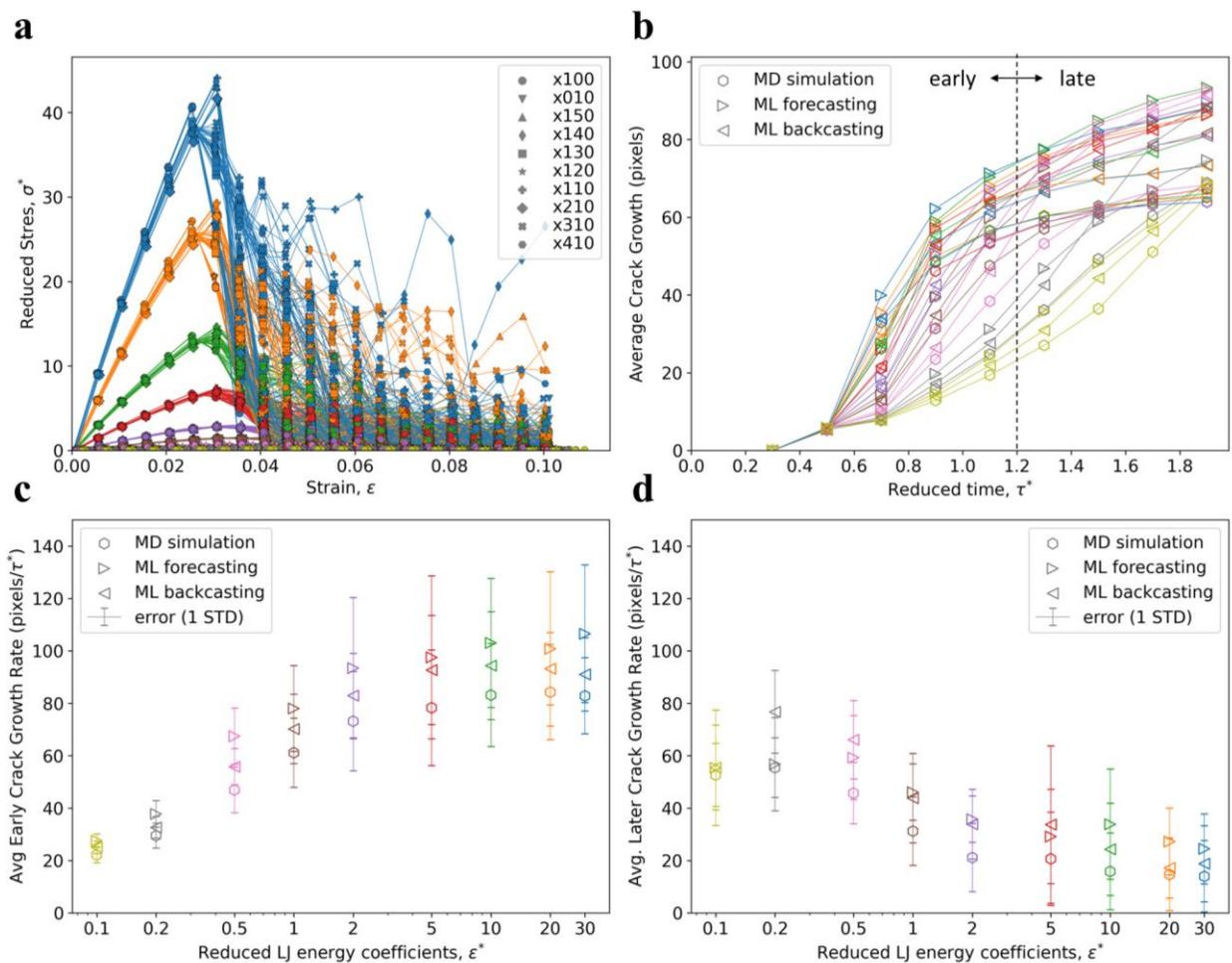

**Fig. 3:** Varied dynamical material responses depending on microstructural changes and alterations of the interatomic potential shape. Panel (a) shows the stress-strain curves from simulations. To evaluate the performance of our predictive model, we compare the results from simulations and predictions by our deep neural network, (b) average crack growth in pixels, and the crack growth rate (c) in the early stages (approximately when $\tau^* < 0.8$) and (d) in the late stages. The colors in panels c and d indicate the energy coefficients for Lennard-Jones potentials (applies to all panels).

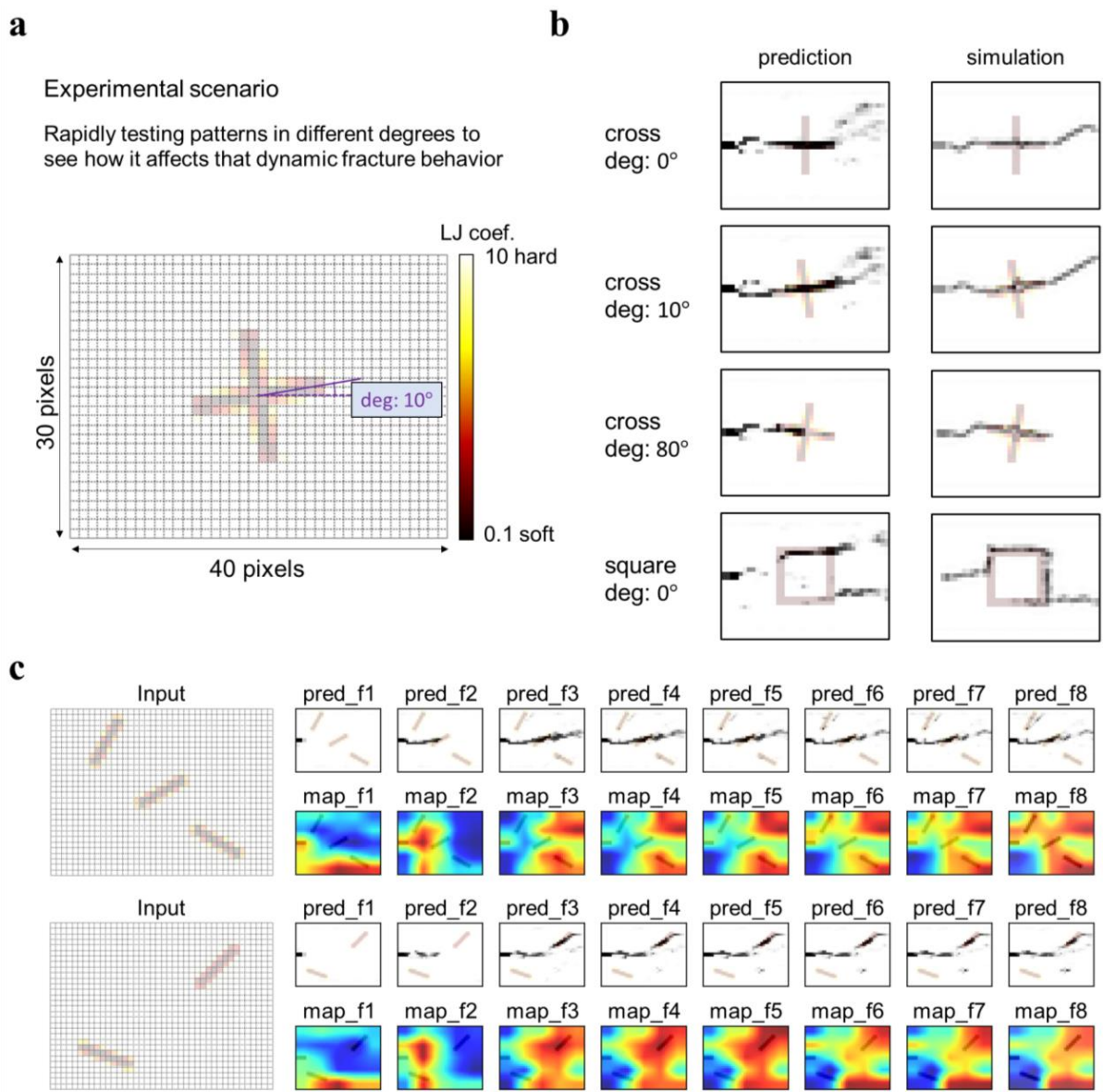

**Fig. 4**: An experimental approach for material design against mode-I failure using our predictive model to forecast the fracture propagation on different geometries. Panel a shows the experimental scenario and the design space we would like to explore, where the LJ coefficients, the key physical factor indicating hard and soft materials are plotted in a hot colormap, ranging from 0.1 to 10 for DyFraNet. Panel b shows the results of this experiment, where only one frame at the certain time point is reported (for the whole history please see **Movie S2**). Panel c shows the activation map, calculated by Grad-CAM method, from the core layer, the layer right after LSTM unit, of DyFraNet. We can see how our model changes its focus from different areas on the input for different crack developing stages. In the initial stage, f1 to f2, the crack should propagate near the initial crack, so DyFraNet focus on the geometries around the crack tip. During the main crack development stage, around f3 to f5, DyFraNet should determine how the crack would occur in what kind of patterns and at what level of pace, which requires a bigger scope to take a wide range of geometries into consideration. Finally, at the final state, f6 to f8, the main crack propagation settles, so the model senses when to stop making prediction of other crack growth and finds the reasonable crack path that links all the defective areas.

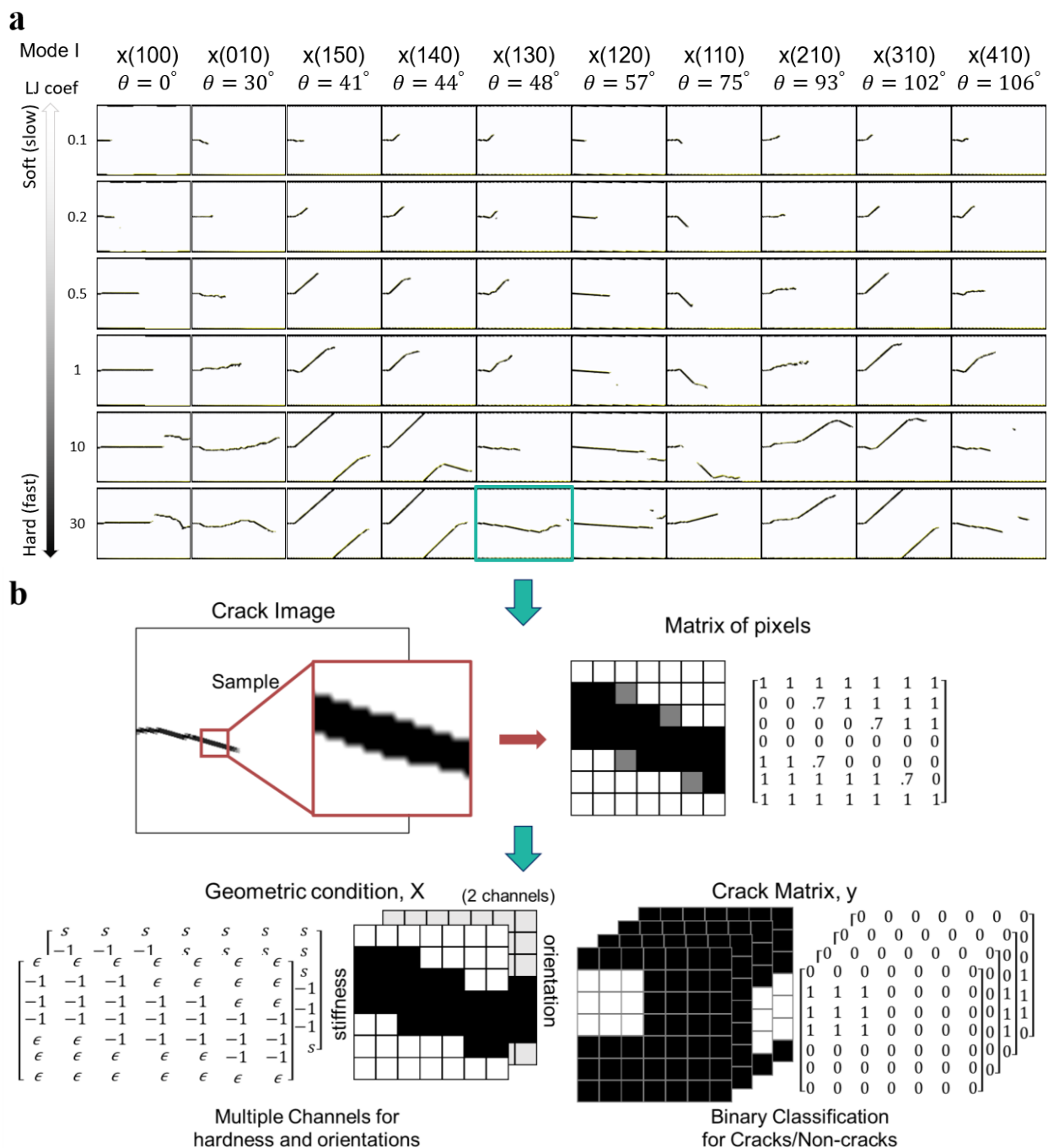

**Fig. 5:** Introduction of our dataset and data representation method. Panel a shows the results of molecular dynamics simulations on crystalline solids for mode I using LAMMPS. Please see **Movie S3** for the movies since here shows only the final timesteps. Panel b shows our image processing techniques to transform these simulation results into matrix form that can be fed into our machine learning model. Please note that for the output crack matrix, y, we use 1 to denote where the cracks are and 0 for non-crack regions. Also, for the input geometric matrix, X, we concatenate two different geometric conditions, one for the orientation, $s$, and the other for the Lennard-Jones energy coefficient, $\epsilon$, by which we can tell the model which target properties to learn to understand the relationship between the target properties and the corresponding fracture behaviors.

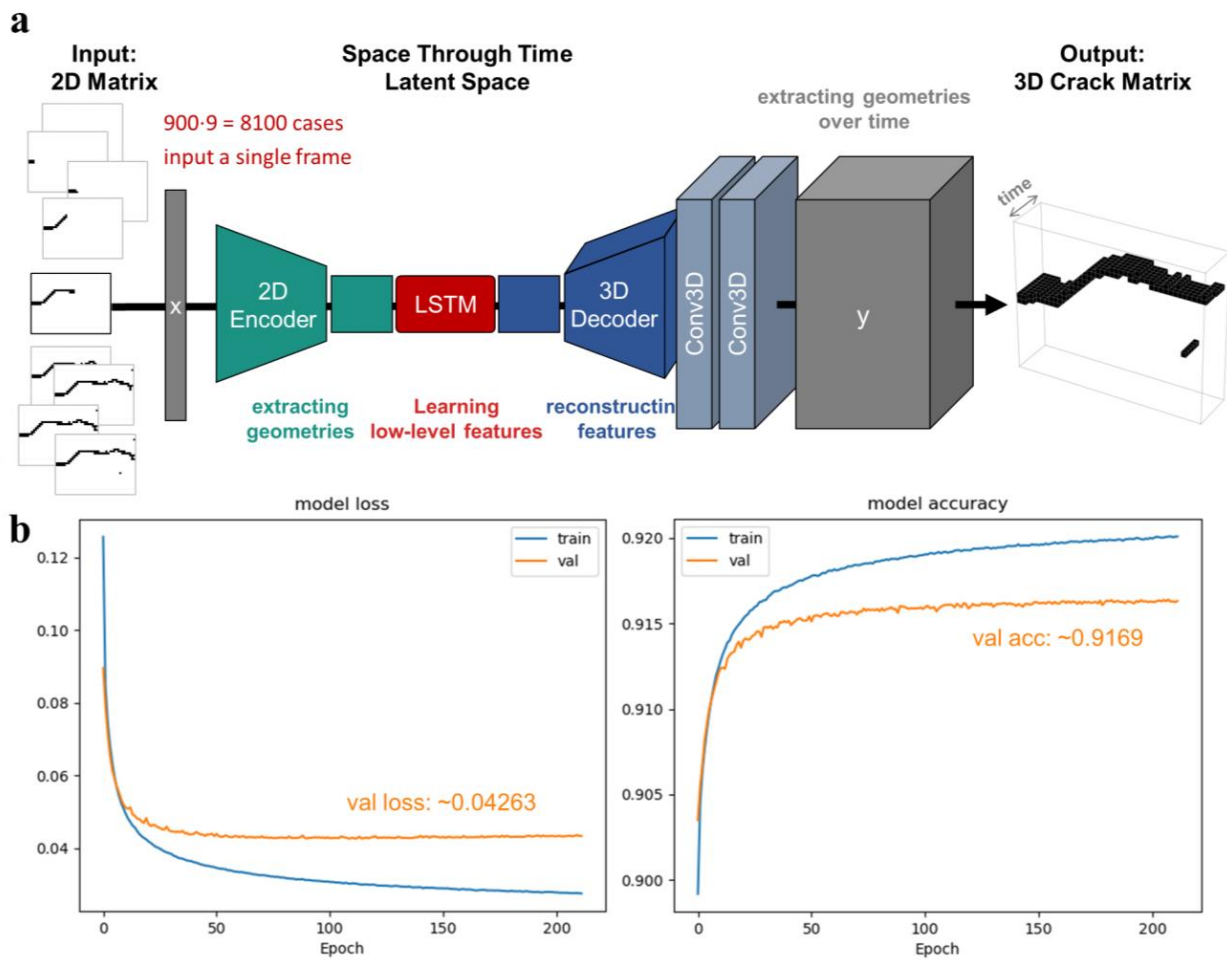

**Fig. 6:** (a) Summary of the model architecture adopted in our work, and (b) the corresponding training history. The model first takes one of the frames as the input into a 2-D encoder to embed geometric features into a latent space. Then, the LSTM unit, which has been proven effective to learning the causal relationship, is going to learn the low-level features in the latent space and translate them into another form, prepared for a 3-D decoder. Finally, the decoder will expand those features into a 3-D crack matrix, the predicted result of fracture propagation. The black voxels identify where the cracks are. The strategy of this approach is to utilize the latent space across a LSTM unit that can introduce information of time, another dimension, into these low-level features, so that the following part of our network can understand certain geometric conditions will lead to what dynamic fracture behaviors over time.